\documentstyle[psfig,epsf]{aipproc}
\input psfig.tex
\begin{document}
\title{Jets, Disks and Spectral States of Black Holes}

\author{Sandip K. Chakrabarti\thanks{Also, Honorary Scientist, Centre for Space Physics, 
IA-212, Salt Lake, Calcutta 700097}}
\address{S.N. Bose National Centre for Basic Sciences\\
JD Block, Salt Lake, Calcutta 700098\\}

\maketitle

\begin{abstract}
We show that outflow rates in jets directly depend on the spectral states of
black holes. In particular, in soft states, when the Comptonized electrons are cold, 
outflow rate is close to zero. In hard states, outflow could be steady, but the rate 
may be very small -- only a few percent of the inflow. In the intermediate states, 
on the other hand, the outflow rate is the highest -- roughly thirty percent of the 
inflow. In this case, piled up matter below the sonic surface of the outflow could become 
optically thick and radiative processes could periodically cool the outflow and 
produce very interesting effects including transitions between burst (high-count 
or On) and quiescence (low-count or Off) states such as those observed in GRS 1915+105.
\end{abstract}

\noindent To appear in the Heidelberg International  Gamma-Ray Astronomy Conference.
Eds. F. Aharonian and H. Voelk (in press).

\section*{Introduction}

Jets from a black hole candidate must form out of accretion flows 
and the surrounding corona since black holes
have no atmosphere of their own. However,
though  black holes have no hard surfaces, they can have `boundary layers'. 
Because of rapid infall near the horizon, viscosity finds little 
time to transport angular momentum from the matter 
and the specific angular momentum becomes almost 
constant close to the horizon. This consideration led earlier workers 
to study thick accretion disks~\cite{pw80}.
The centrifugal force for a constant angular 
momentum flow varies as $\sim 1/r^3$ while the gravitational force
varies roughly as $\sim 1/r^2$. Matter is thus piled up behind 
the strong centrifugal barrier and an oblique shock is formed~\cite{skc96,mrc96}.
At this shock, flow converts kinetic energy into thermal energy
and hard X-rays are emitted from the post-shock region. This
region may thus be called a centrifugal pressure supported boundary 
layer (CENBOL) for all practical purposes.

Post-shock region has similar properties as that of a thick disk,
since the flow puffs up and also has a funnel which can collimate
outflows. In the absence of Comptonising 
soft photons from a pre-shock Keplerian disk (i.e., when the
Keplerian rate ${\dot M}_d$~\cite{ct95} is very low) the post-shock
region remains hot and the spectrum is hard. However, when 
the Keplerian rate is high, the post-shock region cools catastrophically and the CENBOL collapses. 
The emitted spectrum is soft. Chakrabarti \& Titarchuk~\cite{ct95}
and Chakrabarti~\cite{skc97} computed spectra of such types of flows
with or without centrifugal barrier for various parameters. 

Given that the post-shock region behaves like a boundary layer, whose 
electron and ion temperatures depend on the Keplerian and sub-Keplerian accretion rates
(which in turn determine the spectral state of a black hole), it is curious to ask if this boundary 
layer is capable of producing outflows and how the outflow rate depends on spectral 
states. In the next Section, we describe recent results~\cite{yati,caa99,skcpes,skcsam}.

\begin{figure}
\vbox{
\vskip -2.5cm
\hskip -0.0cm
\centerline{
\psfig{figure=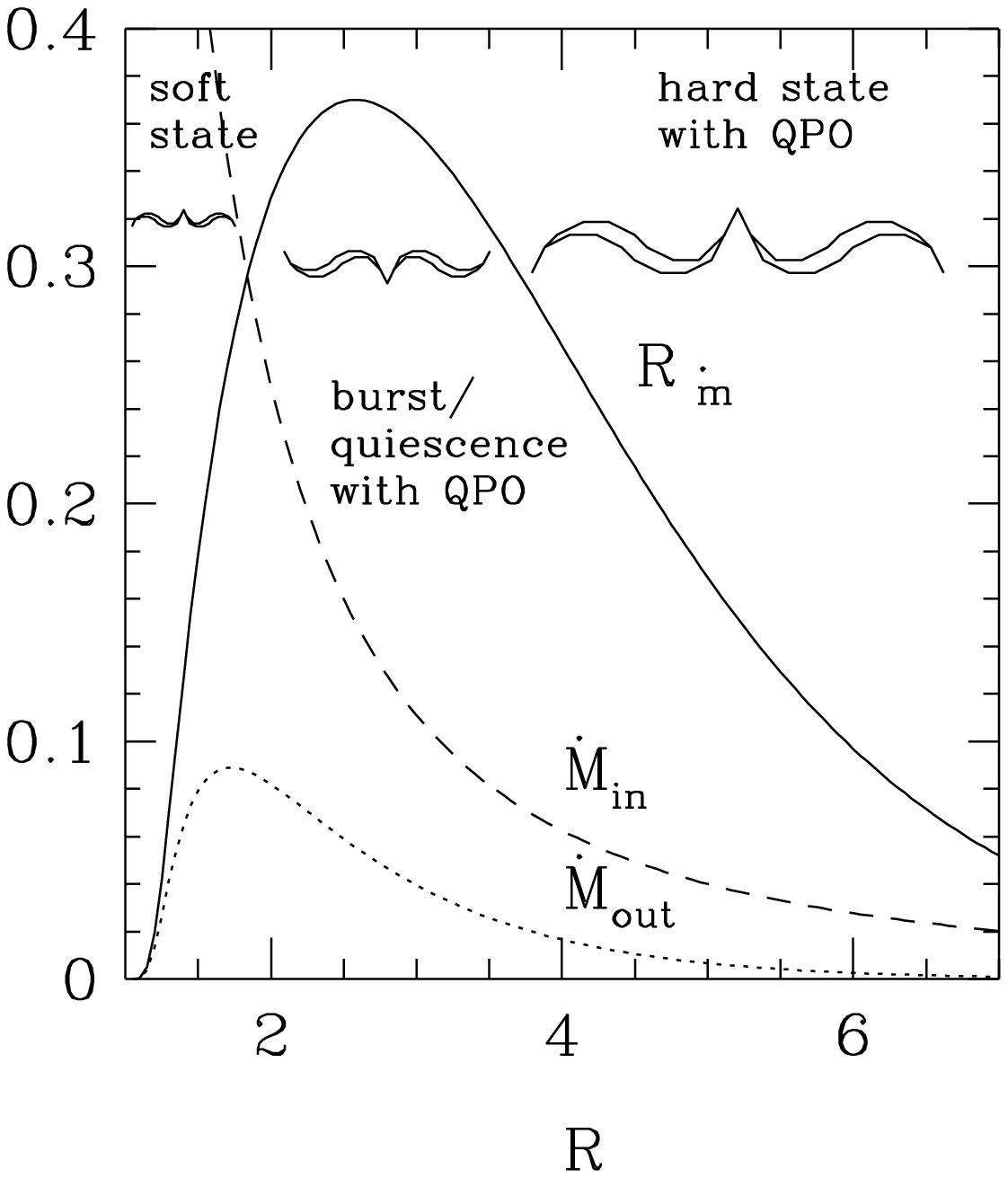,height=10truecm,width=10truecm}}}
\begin{verse}
\vspace{-1.0cm}
\noindent{\small{\bf Fig.1}: Variation of ratio of outflow and inflow rates
as a function of the compression ratio at the accretion shock. Also shown are 
expected outflow rate (short-dashed curve) for a given inflow rate (long-dashed curve).} 
\end{verse}
\end{figure}

\begin{figure}
\vbox{
\vskip 0.0cm
\hskip 2.0cm
\psfig{figure=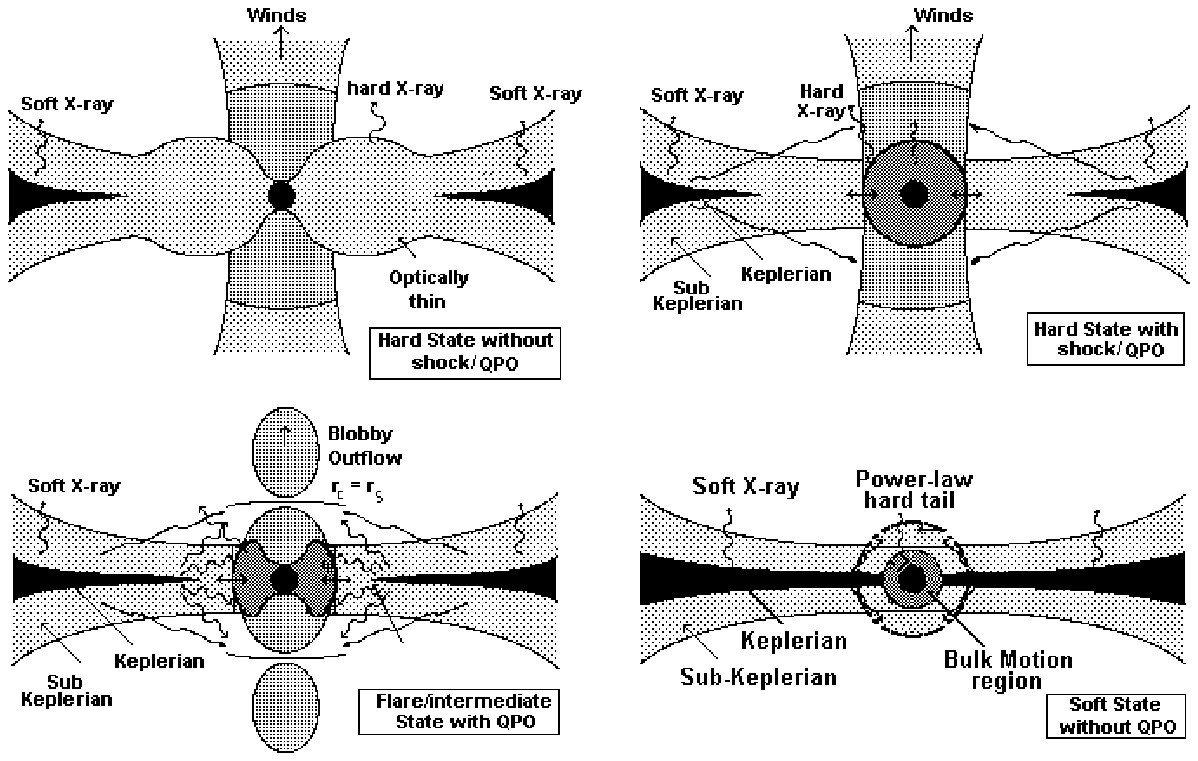,height=8truecm,width=12truecm,angle=90}}
\begin{verse}
\vspace{0.0cm}
\noindent{\small{\bf Fig.2}: Four typical ways accretion/jet system may manifest themselves
within the framework of the advective disk paradigm. Upper left: hard state without QPO;
upper right: hard state with QPO; lower left: Intermediate state with large outflows and
lower right: soft state.  }
\end{verse}
\end{figure}

\section*{Spectral States and Outflow rates from Global Solutions}

Discussion on global solutions has been made in several
recent reviews~\cite{skcpes,gut,skchd,ijp98a} and will not be repeated here.
A hot CENBOL can drive outflows and Chakrabarti~\cite{caa99,ijp98a}
computed the ratio $R_{\dot m}$ of the 
outflow rate (${\dot M}_{out}$) and inflow rate (${\dot M}_{in}$)
as a function of the compression ratio $R$
and the ratio $R_{\Theta}$ of the solid angles subtended by the outflow and
inflow. Figure 1 shows how $R_{\dot m}$ looks like when $R_{\Theta} \sim 1$
is chosen (solid curve). Since in soft states, shocks disappear
($R\rightarrow 1$, left hand side of the box)
and accretion rate is around ${\dot M}_{Edd}$ while in hard states shocks
are the strongest ($R\rightarrow 4-7$, right hand side of the box)
and the accretion rate is low (say, around $\sim 0.001-0.01{\dot M}_{Edd}$, 
see, Chakrabarti \& Titarchuk~\cite{ct95}) we plot 
a `reasonable' ${\dot M}_{in} \propto 1/R^2$ as the long dashed curve.
The product ${\dot M}_{out}={\dot M}_{in} R_{\dot m}$ is plotted as a 
short-dashed curve. It is clear that the maximum outflow rate is
possible when $R\sim 2$, which is possible neither 
in the soft state nor in the hard state. This region 
may be called the intermediate state. Because outflow 
rates are very high, electrons till the sonic sphere
could be cooled down due to Comptonization. A part of the matter
will fall back and the rest would separate as blobs. 
Flow separation takes place at the new sonic surface. Thus,
outflows are expected to be blobby in this intermediate state.
The return flow determines the duration of the On and Off
states. As $R$ approaches $4-7$ from this state, the duration
should progressively increase, provided the geometric
properties of the inflow and outflow (i.e., $R_\Theta$)
remains the same. In hard states, 
continuous outflow would be possible while in
true soft states ($R\sim 1$), outflow would be negligible.

Figure 2 shows various types of accretion disk/jet systems
when the global solutions of advective disks
and radiative transfer effects described above are simultaneously
taken into account. The upper left panel shows the situation
when Keplerian rate is low ${\dot M}_K \sim 0.01-0.001 \ {\dot M}_{Edd}$, 
sub-Keplerian rate is comparatively high and the CENBOL
is not well defined (shock condition not satisfied). Das \&
Chakrabarti~\cite{dc99} showed that the region of the
{\it pressure maximum} may still be considered as a CENBOL and
outflow rates could be computed. Outflow, at a low rate, would be possible.
In this case, shock oscillation does not take place 
and no quasi-periodic oscillations (QPO) are seen.
The upper right panel shows another type of hard state when a 
shock could form and it could oscillate to produce quasi-periodic
oscillations~\cite{msc96,rcm97}. The configuration shown in
the lower left panel is possible in the burst or intermediate (may
also be termed as burst/quiescence or On/Off) states. Here the
light curve can have interesting behavior as is seen in
GRS1915+105. For the On/Off transitions to take place,
it may be crucial that the outflow below the sonic surface
becomes optically thick to Compton scattering in a short time (tens to hundreds of seconds).
If the accretion rate is very small, this never happens and
the light curve does not show On/Off transitions.
The lower right panel shows the situation
when the Keplerian rate is high and the sub-Keplerian component
is weak. CENBOL never forms as it cools down catastrophically.
No wind is produced as well.

Our understanding of the relation between spectral states and outflows 
may have been borne out observationally~\cite{f99,d00}.
It may be interesting to see if the radio jets around super-massive black holes also
show similar behavior. On the other hand, since density in the
wind falls off faster than $1/r$, $\tau$ may never cross unity
for supermassive black holes. Also, presence of magnetic fields
may cool electrons more rapidly and bring interesting 
observational effect. These aspects  would be studied in the future.

This work is partly supported by DST project Analytical and Numerical
Studies of Astrophysical Flows Around Compact Objects.

\end{document}